\documentclass[aps,preprint,preprintnumbers,nofootinbib,showpacs]{revtex4}

\usepackage{amsmath}
\usepackage{amssymb}
\usepackage{mathtools}
\usepackage{dsfont}
\usepackage{color}
\usepackage{graphicx,accents}
\usepackage[normalem]{ulem}

\definecolor{indred}{rgb}{0.8, 0.36, 0.36}
\def\bea{\begin{eqnarray}}
\def\eea{\end{eqnarray}}
\def\sea{\nonumber \\&&}
\def\lla{\left\langle}
\def\rra{\right\rangle}

\def\zc{\gamma}

\def\zb{\beta}

\def\ssc{\scriptscriptstyle}
\def\lsim{\mathrel{\raise.3ex\hbox{$<$\kern-.75em\lower1ex\hbox{$\sim$}}} }
\def\gsim{\mathrel{\raise.3ex\hbox{$>$\kern-.75em\lower1ex\hbox{$\sim$}}} }
\newcommand{\bra}[1]{\lla#1\right|}
\newcommand{\ket}[1]{\left|#1\rra}
\newcommand{\pdv}[2]{\frac{\partial#1}{\partial#2}}

\makeatletter
\DeclareRobustCommand{\cev}[1]{%
  \mathpalette\do@cev{#1}%
}
\newcommand{\do@cev}[2]{%
  \fix@cev{#1}{+}%
  \reflectbox{$\m@th#1\vec{\reflectbox{$\fix@cev{#1}{-}\m@th#1#2\fix@cev{#1}{+}$}}$}%
  \fix@cev{#1}{-}%
}
\newcommand{\fix@cev}[2]{%
  \ifx#1\displaystyle
    \mkern#23mu
  \else
    \ifx#1\textstyle
      \mkern#23mu
    \else
      \ifx#1\scriptstyle
        \mkern#22mu
      \else
        \mkern#22mu
      \fi
    \fi
  \fi
}

\makeatother
\begin{document}
\preprint{{\vbox{\hbox{NCU-HEP-k101}
\hbox{Aug 2023}
\hbox{ed Oct 2023}
}}}
\vspace*{.7in}


\title{\boldmath  Equivalence Principle for Quantum Mechanics in the Heisenberg Picture
\vspace*{.2in}}

\author{Otto C. W. Kong}

\affiliation{ Department of Physics and Center for High Energy and High Field Physics,
National Central University, Chung-li, Taiwan 32054  
\vspace*{.7in}}

\begin{abstract}
We present an exact quantum observable analog of the weak equivalence principle 
for a `relativistic' quantum particle. The quantum geodesic equations are obtained 
from Heisenberg equations of motion as an exact analog of a fully covariant classical
Hamiltonian evolution picture, with the proper identification of the canonical momentum
variables as $p_\mu$, rather than $p^\mu$. We discuss the meaning of the equations 
in relation to projective measurements as well as equations with solution curves as
ones in the noncommutative geometric picture of spacetime, and a plausible approach
to quantum gravity as a theory about quantum observables as physical quantities
including the notion of quantum coordinate transformation.  
\\[.2in]
\end{abstract}

\maketitle

\section{Introduction}
The question of if there is or what could be an equivalence principle (EP)
for quantum mechanics is obviously an important one at the trailhead of our
exploration of the theory of quantum gravity \cite{PPV}. There has been 
a lot of work on the subject matter, at least since the paper by Greenberger 
in 1968 \cite{G}. Summaries of many available results and good lists of 
references are available in Refs.\cite{ah,PPV}.
Here, we are focusing on the dynamics of a quantum particle, hence the weak
version of the principle. Contrary to most, if not all, of the results in
the literature, we are going to give an exact analog of the classical 
picture. That is to say, we have the exact weak equivalence principle (WEP), 
that a quantum particle moves along a quantum geodesic independent of its 
mass, there is local equivalence between gravity and acceleration for it,
and its inertial mass agrees with its gravitational mass. That is to be 
obtained in the Heisenberg picture of quantum dynamics, in an exact 
`relativistic' setting. 

Heisenberg picture analysis of the subject matter has, apparently, hardly 
been performed. For the Schr\"odinger picture analyses, basically, there is 
a kind of consent that the exact weak EP as we have in classical physics 
has to be compromised in some way. A naive reasoning is that a quantum
particle cannot have a definite path of motion, {\em in a classical geometric model 
of space(time)to be exact, hence cannot follow a geodesic in the latter}. And of 
course, the Schr\"odinger picture and the Heisenberg picture are equivalent
descriptions of the same dynamics. The answer to the apparent mystery is 
a fundamentally different perspective as in our term of a quantum geodesic,
instead of `quantum corrected geodesic' \cite{DM} or effective geodesic
 equation \cite{PPV}.
 
A geodesic equation is a differential equation of a distance or length 
parameter that has the shortest path as the solution. Physically, it is the 
equation of motion for a free particle. The equation, of course, governs 
how the position observables change with the motion. Our quantum geodesic 
is exactly such a differential equation for the quantum position observable, 
and that is independent of the state. It is a Heisenberg equation of motion 
for a free quantum particle. To think about the simplest `nonrelativistic' case, 
we certainly have a motion of constant momentum as quantum observables, 
{\em i.e.} $\frac{d\hat{p}^i}{dt}=0$. Note that a conservation law of this
kind is actually exact and of no less importance than its classical analog. 
It is common to read statements that the uncertainty principle says that 
conservation laws are compromised in the quantum setting. Such statements 
are terribly misleading, if not completely wrong. A conservation statement
such as $\frac{d\hat{p}^i}{dt}=0$ certainly cannot give you single constant
eigenvalue answers in projective measurement for any particular $\hat{p}^i$, 
so long as you are not working on an eigenstate of the observable. Yet, the
conservation statement says a lot about the time-independent properties of
the momentum. Not only that the expectation values are time-independent, 
but every physical property or related mathematical result dependent only 
on $\hat{p}^i$ are not changing with time. With projective measurements, 
all the statistical distributions of eigenvalue results obtained for any time
instance (precisely any fixed time after the preparation of the states) for 
any observable as a function of only the three $\hat{p}^i$ would not change
with time. Each state, or ensemble of the same state, would give different
constant distributions. But the constant behavior is a result as important
and as exact as its classical counterpart. The time-independent nature of
any such statistical distribution of results of projective measurements 
can be experimentally verified to any required precision in principle. 
Likewise, a Heisenberg equation of motion is a prediction from the quantum
theory that can be verified precisely free of any concern about quantum
uncertainty. Physics is about physical quantities, {\em i.e.} observables,
and their behavior. Our analysis here hence offers an alternative approach
to look at quantum physics in the presence of gravity that may open a new
path towards quantum gravity. Within practical quantum physics, it could
offer useful results complimentary to the Schr\"odinger picture ones. 

We used the `nonrelativistic' setting to clarify the key background perspective
above for a good reason. Our preferred `relativistic' theory for dynamics is one 
with an invariant evolution parameter $s$ in the place of Newtonian time \cite{096}. 
For all the classical analyses here, the solution gives the particle proper time as 
a linear function of the $s$ parameter, hence essentially identifying the two
physically. It is important to note that we are talking about a Hamiltonian 
dynamical theory with genuinely four degrees of freedom, instead of three as in 
a theory assuming the proper time to be the evolution parameter resulting in the 
velocity constraint. There is no {\em a priori} assumption about a relation between
$s$ and any dynamical variable. The geodesic problem in any manifold as a
variational problem is, of course, one with as many degrees of freedom as the 
dimension of the manifold. In the quantum case, all coordinates and the particle
proper time should be seen as quantum observables, or operators, while $s$
stays as a real parameter characterizing the Hamiltonian evolution. Such a 
description of quantum dynamics, while available since at least around 1940 
 \cite{Fo,S} (see for example the books of Refs.\cite{H,F} and references therein), 
is not what is commonly presented in textbooks. Even the `standard' presentation 
of classical Hamiltonian dynamics does not do that (see however Johns\cite{J}) 
and hence does not give a Lorentz covariant formulation \cite{096}.  Our results 
here also illustrate an advantage of that formulation of  `relativistic'  dynamics, 
classical and quantum.
 
Our analysis starts in the next section with a presentation of the classical
picture, especially focusing on a formulation in terms of Hamiltonian dynamics
with the phase space seen as a cotangent bundle of the spacetime as the 
configuration space of the particle. The quantum dynamics in the Heisenberg
picture and the quantum geodesic equation are then  straightforward to 
obtain along the line. That is presented in section \ref{sec3}. The last section 
gives careful discussions of various related aspects of the theory of quantum
mechanics. While the Heisenberg picture analysis is not dependent on the
explicit theory for the corresponding, even abstract vector space, Schr\"odinger 
picture, our new theory of `relativistic' quantum mechanics with a notion of
Minkowski metric operator for the vector space of states \cite{096,087}, 
defining the inner product, is conceptually deeply connected to our Heisenberg 
picture results. In particular, we emphasize a perspective that takes seriously 
the quantum observables as physical quantities to be understood beyond 
the framework of classical physical concepts. In particular, the position and 
momentum observables may be seen as coordinate observables of the 
quantum phase space  \cite{078} as a noncommutative geometry \cite{C,Ma}. 
Some projections on taking an approach to quantum gravity focusing on the
dynamical behavior of the physical quantities as quantum observables/operators
are also discussed, including the important notion of quantum coordinate
transformations in relation \cite{Ha}.
 
\section{Geodesic and Hamiltonian Dynamics -- Classical Case}
For the background analysis used in the section, we follow the presentation
in the lecture note by Tong\cite{T}. The latter gives a careful derivation of the
geodesic equation through minimizing the action
\bea
S_o = \int\!\!ds L_o(s) = \int\!\!ds \sqrt{ -  g_{\mu\nu}(x) \frac{dx^\mu}{ds} \frac{dx^\nu}{ds} } \;,
\eea
that we are not repeating here. Clearly, one would obtain the same equation
taking the Lagrangian $L(s)= -\frac{m}{2} L_o^2$. The latter, called 
a `useful trick' by Tong, is the exact one that gives, through a Legendre 
transformation, a covariant Hamiltonian formulation of the free particle 
dynamics that is the exact `relativistic' extension of the `nonrelativistic'
Hamiltonian formulation. The Lagrangian and the Hamiltonian,
$H(s)= p_\mu \dot{x}^\mu - L(s)$ are both just the kinetic term 
$\frac{m}{2} g_{\mu\nu}(x) \frac{dx^\mu}{ds} \frac{dx^\nu}{ds}
  =  \frac{1}{2m} g^{\mu\nu}(x) p_\mu p_\nu$, where the canonical momentum
variables are $p_\mu= \pdv{L}{\dot{x}^\mu}$, $\dot{x}^\mu \equiv \frac{dx^\mu}{ds}$,
instead of $p^\mu$. Hamilton's equations of motion, $\frac{dx^\mu}{ds} = \pdv{H}{p_\mu}$
and $\frac{dp_\mu}{ds} = -\pdv{H}{x^\mu}$, are special examples of equation
of motion for all observables $F(x^\mu,p_\nu)$ given by
\bea
\frac{dF}{ds} = \{F,H\}= \pdv{H}{p_\mu}\pdv{F}{x^\mu}-\pdv{F}{p_\mu}\pdv{H}{x^\mu} \;,
\eea
and the canonical condition for the coordinates of the covariant phase space is
given through the Poisson brackets
\bea
\{ x^\mu, x^\nu \}= \{p_\mu, p_\nu\}=0 \;,
\qquad
\{ x^\mu, p_\nu \} = \delta^\mu_\nu  \;.
\eea
Note that in the presence of a nontrivial $g_{\mu\nu}$, $x^\mu$ are not 
components of a four-vector, but $p_\mu$ are. We avoid writing $x_\mu$ here but 
$p^\mu = g^{\mu\nu} p_\nu$ are well-defined. Yet, even for canonical $p_\mu$,
while one has $\{ x^\mu, p^\nu \} = g^{\mu\nu}$, $\{ p^\mu, p^\nu \}$ are generally
nonzero. For a generic Riemannian manifold with $x^\mu$ as local coordinates, 
the canonical momentum is still a cotangent vector and the phase space as the 
cotangent bundle is a symplectic manifold  by construction. The geometry of the 
latter dictates the structure of the local Hamiltonian dynamics. $s$ as the parameter 
of the Hamiltonian flows has its mathematical nature fixed by the choice of 
Hamiltonian. For the case at hand, one can easily see that it is indeed essentially 
a  geodesic length parameter and the physical proper time of the particle. For 
general use of configuration and momentum variables, we have the Poisson 
bracket given by
\bea
 \{F,G\} =  {\mathcal P}^{\mu\nu} \pdv{F}{x^\mu}   \pdv{G}{x^\nu} 
 + {\mathcal P}^{{\mu}}_{\tilde{\nu}} \left( \pdv{F}{x^\mu}   \pdv{G}{p_{\nu}} 
    -\pdv{G}{x^\mu}   \pdv{F}{p_\nu} \right)
 + {\mathcal P}_{\tilde{\mu}\tilde{\nu}}  \pdv{F}{p_{{\mu}} }   \pdv{G}{p_{{\nu}} } \;,
\eea
with the canonical condition given by
\bea
{\mathcal P}^{\mu\nu}  = {\mathcal P}_{\tilde{\mu}\tilde{\nu}} =0 \;,
\qquad 
{\mathcal P}^{\mu}_{\tilde{\nu}} = \delta^{\mu}_{\nu} \;,
\eea
with indices $\mu$ and $\nu$ referring to the position coordinates and
$\tilde{\mu}$ and $\tilde{\nu}$ referring to the corresponding momentum
coordinates.  

Let us go on adopting Tong's illustration of the WEP with the supplement of 
the Hamiltonian picture. We consider particle dynamics under (constant) gravity 
in a Minkowski spacetime and its description in the instantaneous frame of 
a free-falling observer (with the Kottler-M\"oller coordinates, at $\rho(\tau)=0$). 
In terms of the classical metric
\bea
ds^2 
=-c^2dt^2+dx^2+dy^2+dz^2
= -\left(1+\frac{a\rho}{c^2} \right)^2 c^2 d\tau^2 +d\rho^2 +dy^2+dz^2\;,
\eea
with
\bea\label{cct}
ct = \left(\rho +\frac{c^2}{a} \right) \sinh\!\left(\frac{a\tau}{c} \right)\;,
\qquad
x = \left(\rho +\frac{c^2}{a} \right) \cosh\!\left(\frac{a\tau}{c} \right) \;. 
\eea
$a=\sqrt{-(a^t)^2+ (a^x)^2}$, $a^x= \frac{d^2x}{d\tau^2} = a  \cosh\!\left(\frac{a\tau}{c} \right)$
and $a^t = \frac{d^2ct}{d\tau^2} = a  \sinh\!\left(\frac{a\tau}{c} \right)$,
is the constant acceleration, as the value of  $\frac{d^2x}{dt^2}$ at the
observer's rest frame.

A  Hamiltonian for the classical dynamics of the constant acceleration,
under the original frame of reference, can be given by
\bea \label{h}
{H}_{\!a}(s) = \eta^{\mu\nu} \frac{p_\mu p_\nu}{2m} - \frac{ma^2}{2c^2}\eta_{\mu\nu}x^\mu x^\nu \;,
\eea
with $\eta_{\mu\nu}=\mbox{diag}\{-1,1,1,1\}$ as the Minkowski metric.
Note that the simple kinetic term guarantees $p^\mu=  m \frac{dx^\mu}{ds}$, 
which says $s$ is the particle's proper time. The variable $\tau$ above is 
taken as the time coordinate of the instantaneous free-falling frame, hence 
differs from $s$ in general. That is to say, other than the special solution of 
Eq.(\ref{cct}), we do not have $\tau=s$. A key question is if the transformation 
should be seen as a canonical one. We are taking the configuration space 
coordinate transformation $x \to x'$ onto the phase  space by enforcing the 
momentum four-vector to transform as dictated by that, which preserves the
metric independent result $p'^\mu=  m \frac{dx'^\mu}{ds}$. Explicitly, we have
\bea\label{pt}
p^{ct} = \pdv{ct}{\rho} p^\rho + \pdv{ct}{c\tau} p^{c\tau}\;,
\qquad
p^x = \pdv{x}{\rho} p^\rho  + \pdv{x}{c\tau} p^{c\tau} \;.
\eea
That is to say, one simply takes the coordinate transformation on the 
configuration manifold to its cotangent bundle. Conceptually, the $p'_\mu$, 
from $p'_\mu dx'^\mu = p_\nu dx^\nu$, are still the canonical components 
of the cotangent vector. One can confirm that analytically by checking the 
canonical condition through evaluating the Poisson brackets among the
new canonical position and momentum variables $(c\tau,\rho,y,z)$ and
$(p_{c\tau}, p_\rho ,p_y,p_z)$. The kinetic term of ${H}_{\!a}(s)$ maintains 
the quadratic form, as $g^{\mu\nu}(x')\frac{p'_\mu p'_\nu}{2m}$, while the 
potential term, neglecting the $y$ and $z$ dependent part, is 
$\frac{-m(c^2+a\rho)^2}{2c^2}$, giving free particle motion 
instantaneously at for $\rho=0$ at $s=0$. Explicitly, with the initial values 
at $s = 0$ for all original phase space coordinate variables being zero except 
for $x = \frac{c^2}{a}$ and $p^{ct} = mc$, one obtains the solution of 
Eq.(\ref{cct}). We have $\tau = s$ and $\rho$ and $p^\rho$ maintaining their 
vanishing values, while $p^{c\tau} = mc$, and $H_{\!a}$ has value of $mc^2$.

Complete results of the Hamiltonian dynamics above in the new and old coordinates,
as well as the exact geodesic equations under the $g_{\mu\nu}$ metric can easily  
be worked out. We refrain from giving them explicitly here but their exact quantum
analogs are to be given below. We want to note that in the Hamiltonian $H_{\!a}$,
the mass parameter $m$ in the kinetic term is really the inertial mass, as in
$p^\mu=m\frac{dx^\mu}{ds}$, while the one in the potential term is a gravitational
one. Without the equality of the two masses, there is no mass-independent free fall
as obtained.
 
\section{Geodesic and Hamiltonian Dynamics -- Quantum Case \label{sec3}} 
Now in this section, we are going to illustrate the exact quantum analog of the 
classical analysis above with the classical observables replaced by quantum 
observables. The Hamiltonian approach can be used as a useful guideline, and
results can most easily be seen from the Schr\"odinger representation with
$\hat{x}'^\mu = {x}'^\mu$ and $\hat{p}'_\mu = -i\frac{\partial~}{\partial x'^\mu}$, 
taking $\hbar=1$, but is only a consequence of the commutation relation among 
the operators, or rather just as abstract quantum observables. The commutation
relations are essentially Poisson bracket relations. In fact, we will use 
the terms  operators below, without really committing to any concretely given 
vector space they have to act on. If one prefers to think about that, it is certainly 
fine, so long as it gives a consistent representation of the algebra involved. 
We are more interested in working in the free-falling frame, with the nontrivial 
$g_{\mu\nu}$. The exact quantum Poisson bracket is $-i[\cdot,\cdot]$, giving
the canonical condition as
\bea\label{ccr}
-i[\hat{x}'^\mu, \hat{x}'^\nu] = -i [\hat{p}'_\mu, \hat{p}'_\nu] = 0 \;,
\qquad
-i[ \hat{x}'^\mu, \hat{p}'_\nu] = \delta^\mu_\nu \;,
\eea
which is in fact trivially satisfied under Schr\"odinger representation for any 
choice of the set of position observables. Otherwise, so long as one adopts 
the commutation relations among the $\hat{x}^\mu$ and $\hat{p}_\mu$, for
the case of the Minkowski metric, they can be directly derived by promoting
the classical coordinate transformation to one among the quantum observables 
(which can be seen as noncommutative coordinates of the quantum phase space
 \cite{078}). Explicitly, that means taking Eq.(\ref{cct}) and Eq.(\ref{pt}) as for 
 the operators.

For the Hamiltonian dynamics, we promote $H_{\!a}$ of Eq.(\ref{h}) directly to 
the Hamiltonian operator and write, in the free-falling frame,
\bea\label{Ha}
 \hat{H}_{\!a} =  - \frac{c^4}{(c^2+a\hat\rho)^2} \frac{(\hat{p}_{c\tau} )^2}{2m}
   + \frac{(\hat{p}_\rho)^2}{2m} - \epsilon \frac{m(c^2+a\hat\rho)^2}{2c^2} \;.
\eea
Note that we have dropped the part for the $\hat{y}$ and $\hat{z}$ degrees of 
freedom, for simplicity, and have put in an extra parameter $\epsilon$ for the easy 
tracing of the exact free case with the geodesic in the equations of motion. That is, 
$\epsilon=1$ gives the exact quantum dynamics of what is sketched in the last 
section, while $\epsilon=0$ gives the geodesic motion. The Heisenberg equations 
of motion we are interested in are given by
\bea
\frac{d\hat{p}_{c\tau}}{ds}  = 0 \;,
\qquad
\frac{d\hat{p}^\rho}{ds}  =\frac{d\hat{p}_\rho}{ds} 
=  -\frac{a(c^2+a\hat\rho)}{mc^4} (\hat{p}^{c\tau} )^2 +  \epsilon \frac{ma(c^2+a\hat\rho)}{c^2}  \;,
\eea
and $\hat{p}^{c\tau} = m \frac{d{c\hat\tau}}{ds}$ and $\hat{p}^\rho = m \frac{d\hat\rho}{ds}$.
They have exactly the same form as the classical equations, with the  classical 
observables replaced by quantum ones. We have, however, 
\bea
\frac{d\hat{p}^{c\tau}}{ds}  = -i [\hat{p}^{c\tau}, \hat{H}_{\!a}]
=   - \frac{a}{c^2+a\hat\rho} \frac{\hat{p}^{c\tau}  \hat{p}^\rho}{m}  
  - \frac{ \hat{p}^\rho \hat{p}^{c\tau} }{m}  \frac{a}{c^2+a\hat\rho}
\eea
as the quantum version of the classical geodesic one. Note the sum of the two 
terms on the right-hand side for the classical limit with commutating variables. 
The operator ordering ambiguity from the classical to the quantum case is 
resolved through the Hamiltonian formulation. As in the classical case, the 
second-order differential equations for the position observables are 
mass-independent. Explicitly, they are
\bea&&
\frac{d^2{c\hat\tau}}{ds^2} + \frac{d c\hat\tau} {ds} \frac{a}{c^2+a\hat\rho}    \frac{d \hat{\rho}}{ds} 
+  \frac{d \hat{\rho}}{ds}    \frac{a}{c^2+a\hat\rho} \frac{d c\hat\tau} {ds}=0 \;,
 \sea\label{geo}
\frac{d^2\hat\rho}{ds^2} + \frac{d c\hat\tau} {ds} \frac{a(c^2+a\hat\rho)}{c^4}     \frac{d c\hat\tau} {ds} -\epsilon  \frac{a(c^2+a\hat\rho)}{c^2} = 0 \;.  
\eea
Again, the $\epsilon=0$ case is the free motion, {\em i.e.} quantum geodesic equations.
For the latter, as well as the $\epsilon=1$ case, the equations of motion are 
mass-independent, so long as the inertial mass and the gravitational mass as the
$m$ is the kinetic and potential terms of Eq.(\ref{Ha}), respectively, are taken the same.
All are exact analogs of the classical case.

\section{Discussions}
\subsection{More about the quantum theory}
We have obtained exact quantum analogs of the WEP for the classical case as treated 
in Tong's book supplemented by a fully covariant Hamiltonian dynamical picture based
on an invariant evolution parameter that is essentially the classical particle proper time. 
The Hamiltonian evolution equations, with the clear identification of canonical variables,
are of great help here for getting around otherwise nontrivial operator ordering issues. 
We have emphasized the cotangent bundle structure of the manifold as a phase space 
that gives the right picture of the coordinate transformation for the spacetime, the 
configuration space of the particle, as a canonical coordinate transformation for the 
Hamiltonian dynamics. The quantum part of the story is then based on the adoption 
of position and momentum observables satisfying the canonical condition, Eq.(\ref{ccr}). 
The condition is independent of the (configuration space) metric. After all, symplectic
geometric structure, and hence the Hamiltonian formulation, is independent of even
the very existence of a Riemannian metric, for the configuration space or otherwise. The 
quantum geodesic equation governs the quantum evolution of the position observables, 
which we have argued are exact equations of motion that can be verified. Our result
may well complement those available in the literature and provide a peek into an 
alternative approach to quantum gravity based on the quantum observables. Note 
that only they are the representations of the physical quantities in the theory.  
Spacetime as a physical object should be described with physical quantities, hence 
quantum, instead of classical, observables in a quantum theory. In a theory of particle
dynamics, the only physical notion about spacetime is the observable position
coordinates of a particle. The thinking about the validity of the classical geometric
models of Newtonian space or Minkowski spacetime for the quantum theory hence 
lacks justification. In relation to that, it is interesting to note that the particle proper 
time as the time coordinate in the particle rest frame is really an observable 
($\sqrt{g_{\mu\nu}(\hat{x})\hat{x}^\mu \hat{x}^\nu}$), while the parameter $s$ 
characterizing the Hamiltonian evolution stays 
as a real parameter, as the analog of Newtonian time in the `nonrelativistic' theory. 
The possibility of going beyond that is a fundamental question about the concept
of symmetry of quantum systems as exemplified by the notion of quantum reference 
frame transformations \cite{qrf}. This has very important implications for the EP and 
quantum gravity \cite{Ha}  beyond the classical coordinate transformation analyzed 
here, to be discussed more below. 

Our Heisenberg picture analysis relies only on the Poisson bracket relations among 
the observables and the Hamiltonian formulation of the dynamics among them. It
does not depend on any particular representation picture having the observables 
as explicit operators on a vector space of states, hence not even the notion of 
Hermiticity. The latter is really to be defined based on the chosen inner product 
on the vector space of states. Even the imaginary number $i$ and suppressed 
$\hbar$ in the commutation relations may be unnecessary. One can 
replaced the $-i [\cdot,\cdot]$ by $\{\cdot,\cdot\}_{\!\ssc Q}$ or even just 
$\{\cdot,\cdot\}$ as the (quantum) Poisson bracket., where the exact parallel of 
the classical and the quantum case would be plainly obvious. The triviality in the 
case is because the Hamiltonian analysis for it involves no ordering ambiguity 
going from classical to quantum.

It is well known that Schr\"odinger quantum mechanics is Hamiltonian dynamics.
Hence, the same must be true for the Heisenberg picture description as well. In
fact, one can illustrate clearly that the Heisenberg equation of motion is exactly
a Hamiltonian equation of motion for observables with the Poisson bracket as
identified, essentially already be Dirac. One can take the (projective) Hilbert space 
as the symplectic manifold. Each operator $\zb$ on the Hilbert space can be matched 
to its expectation value function $f_{\!\ssc\zb}=\frac{ \bra\phi \zb\ket\phi}{\lla\phi |\phi\rra}$
as a `classical' observable for which the  two pictures of the Hamiltonian dynamics 
can be matched perfectly with the introduction of a noncommutative (K\"ahler)
product among such function satisfying \cite{cmp,078}
\bea\label{kp}
f_{\!\ssc\zb} \star_{\!\ssc\kappa} f_{\!\ssc\zc} =   f_{\!\ssc\zb\zc}  \;.
\eea
The expression of the Heisenberg equation of motion in terms of the such 
functions for all the observables is the exact Hamilton equation of motion for 
any $f_{\!\ssc\zb}$ in terms of the exact Poisson bracket for the (projective) 
Hilbert space.  The set of the corresponding equations of motion for the real 
or complex number coordinates characterizing the state is the exact content 
of the Schr\"odinger equation. We have suggested the interpretation of the 
Heisenberg picture as the noncommutative coordinate picture of exactly the 
same symplectic geometry and established a consistent differential geometric
picture of that as a noncommutative geometry \cite{078,081}. An exact Lorentz
covariant version of that can be obtained, from a symmetry theoretical
formulation of the `relativistic' quantum picture based on a Lie group/algebra
we called $H_{\!\ssc R}(1,3)$ with Lorentz symmetry plus Minkowski four-vector 
generators $Y_\mu$ and $P_\mu$ together with a central charge $M$ giving 
the above canonical condition/commutation relation  in the form
$[m\hat{x}_\mu, \hat{p}_\nu]= [\hat{y}_\mu, \hat{p}_\nu]=i(\hbar) \eta_{\mu\nu} \hat{m} 
   =i(\hbar) \eta_{\mu\nu} m\hat{I}$,  where $m$ is effectively a Casimir invariant
for the irreducible representation to be interpreted as Newtonian mass \cite{096}. 
An abstract Fock state basis as well as coherent state wavefunction description 
of that has been essentially presented \cite{096,087}. What is particularly 
interesting to note, in relation to the present analysis, is that the theory has
a representation space that is Krein, instead of Hilbert. That is, the inner product 
is not positive definite \cite{Bo}. In fact, it is defined in terms of Pauli's metric
operator \cite{P,D,M} that is for the case Minkowski, denoted by $\hat{\eta}$. 
The position and momentum operators are exactly $\eta$-Hermitian, {\em i.e.}
satisfying
\bea \label{eH}
\prescript{}{_\eta}{\!\lla  \cdot | \zb \cdot \rra}
= \prescript{}{\eta}{\!\lla  \zb \cdot | \cdot \rra} \;.
\eea
One can see that as a special case, the Minkowski case, of a quantum theory
with an inner product defined in terms of a metric operator denoted by $\hat{g}$, 
as $ \prescript{}{_g}{\!\lla  \cdot | \cdot \rra } = \lla \cdot | \hat{g} |\cdot \rra$, or
equivalently $ \prescript{}{_g}{\!\lla \cdot \right|} = \lla \cdot \right| \hat{g}$. The
proper definition of the adjoint or Hermitian conjugate of an operator $\zb$ is
then given by $\zb^{\dag^g}$ satisfying
\bea \label{pH}
\prescript{}{_g}{\!\lla  \cdot | \zb^{\dag^g} \cdot \rra}
= \prescript{}{g}{\!\lla  \zb\cdot | \cdot \rra} \;.
\eea
$g$-Hermitian operators then generate one-parameter groups of 
(pseudo)-unitary transformations that preserve the inner product. 
The usual Hilbert space theory is exactly the case
 for the metric operator being the identity, giving an Euclidean metric on it. 
 The Minkowski nature of such four-vector coordinate observables beyond 
 the naive notion of switching between lower and upper indices is 
 what is needed to justify seeing the position and momentum operators as
 noncommutative coordinates of the quantum phase space (the vector space
 of states or rather its projective space) with a Minkowski metric.
 
 \subsection{On quantum mechanics in nontrivial gravitational background, and beyond}
 The discussion in the last paragraph, apart from giving a solid vector space
 formulation of the `relativistic' quantum dynamics behind the otherwise 
 abstract Heisenberg picture dynamics analyzed above, along but different
 from the line of fully Lorentz covariant formulation \cite{H,F}, also brings up
 an interesting perspective for a plausible picture of a theory of quantum 
 mechanics in a generic curved spacetime and its interpretation as particle
 dynamics on a curved noncommutative geometry. For example, one can think
 about a Schr\"odinger wavefunction representation of the Minkowski picture
 with the integral inner product between $\phi(x)$ and $\psi(x)$ giving by
 \bea
\int\!d^4x \psi_\eta^* (x) \phi(x) =  \int\!d^4x [\hat\eta_{\ssc S} \psi (x)]^*  \phi(x) \;,
\eea
where $\hat\eta_{\ssc S}$ is the explicit representation of $\hat\eta$ as
operator acting on the wavefunction. When we apply the configuration
space coordinate transforming of Eq.(\ref{cct}) and Eq.(\ref{pt}), the $x \to x'$ 
transformation taking the state wavefunctions to $\phi'(x')$ and $\psi'(x')$ 
would take the above integral to
 \bea
\int\!d^4x' \left|\pdv{x}{x'}\right|  [\hat\eta_{\ssc S} \psi'(x')]^*  \phi'(x') 
= \int\!d^4x' \left( 1+ \frac{a\rho}{c^2}\right) [\hat\eta_{\ssc S} \psi'(x')]^*  \phi'(x') \;.
\eea
One can reasonably expect $\hat\eta_{\ssc S}$ to depends on the position 
operators $\hat{x}^\mu$ only, then the inner product integral can be as
\bea
\int\!d^4x' \left( 1+ \frac{a\rho}{c^2}\right) \hat\eta_{\ssc S}(x) \psi'(x')^*  \phi'(x')
= \int\!d^4x'   [\hat{g}_{\ssc S}(x') \psi'(x')]^*  \phi'(x')  \;.
\eea
That is to say, the new description of the quantum theory with states described by
Schr\"odinger wavefunction $\phi(c\tau,\rho,y,z)$ would have a inner product 
defined in terms of the new form of Pauli's metric operator $\hat{g}_{\ssc S}$.
In any case, as the metric (tensor) changes, the inner product changes accordingly,
as what is to be expected from the consistency of the notion of the quantum metric
realized in the form of the inner product on the vector space of states. 

In fact, in our main analysis in section \ref{sec3} above, we have implicitly taken
components of the metric tensor $g_{\mu\nu}$, or its inverse, as functions of
the position observables $\hat{x}'$. In the Schr\"odinger representation, $\hat{x}'$
of course are just $x'$. That is to say, we have already taken the classical 
$g_{\mu\nu}(x)$ to a quantum $\hat{g}_{\mu\nu}= g_{\mu\nu}(\hat{x})$. Our
quantum geodesic equations further illustrate an explicit notion of quantum
Christoffel symbols. Up to operator ordering issues, we have them simply as 
direct operator versions of the classical ones. They can be seen as among the 
class of special elements of the quantum observable algebra that characterize 
the geometry of the quantum picture of the spacetime. While Ref.\cite{PPV} has 
the kind of operators written down, they are not used much in the actual analysis, 
and certainly not from our perspective. The success of the approach here may
be seen, naively, as suggesting an approach to quantum gravity as simple as 
recasting Einstein's theory in operator form. However, other than issues about
the proper operator ordering, there are still important questions to address. We 
name two here. One is the notion of a quantum coordinate transformation we
have touched on, which is to be discussed below. The other is related to Pauli's 
metric operator. The latter gives a metric really on the vector space of states, 
which is generally a K\"ahler manifold, hence the symplectic  structure is tied to 
the metric structure. The corresponding metric tensor is one for the infinite
-dimensional real/complex manifold. The metric tensor operator  $\hat{g}_{\mu\nu}$
is, however, a metric for the configuration space with the position observables,
$\hat{x}'$, or coordinate observables. It is hard to think about that space as
a spacetime any different from the classical one. The answer to the puzzle lies in 
the fact that the quantum phase space, unlike its classical analog, is an irreducible 
representation of the background (relativity) symmetry. In the classical theory 
as an approximation, in which the commutation relations are trivialized, the 
representation is reducible to that of the configuration and momentum space.
That is very much like the relation of the classical Minkowski spacetime to its
`nonrelativistic' approximation as Newtonian space and time. At the `relativistic'
level, there is no separate notion of space and time. At the quantum level, there
is no separate notion of configuration and momentum space. The noncommutative
geometric picture for the quantum phase space is the quantum model of the
physical spacetime. It is not clear then if one should be thinking about only
a metric tensor operator as for the $\hat{x}'$ part only, though that is naturally 
doubled with a copy for the  $\hat{p}'$ part. For example, a generic coordinate
system could have mixed the position and momentum coordinate observables. 
The question is really if quantum gravity is about the metric of the quantum 
phase space as a noncommutative geometry. 

L\"ammerzahl stated that `Quantum mechanics is a non-local description 
of matter' \cite{L}. That statement is probably the first idea many physicists
have in mind when approaching the problems related to the EP. Obviously,
if a quantum particle generally cannot go along an exact path, in a classical
model of space(time), it cannot follow such a geodesic. We have already
discussed much about the idea of our quantum geodesic. With the idea of 
the quantum model of spacetime being a noncommutative geometry, motion 
along a definite path therein is completely feasible. In the Schr\"odinger 
picture, the state of the particle certainly evolves along a definitive path.
Apart from the spacetime geometry picture, quantum nonlocality as in
entanglement between parts of a composite system may be seen as the
deeper meaning of L\"ammerzahl's statement. Yet, even that notion of
nonlocality has been challenged, interestingly enough, in the Heisenberg 
picture  \cite{DH}. Ways to describe the complete quantum information 
for a composite system as information about a set of local basic observables, 
such as the position and momentum of the individual particles have been
presented \cite{097}. For our result of the quantum geodesic, as equations 
of motion for the observables, they are state-independent. One can think 
about two quantum particles in simultaneous free fall. To the extent that 
we can neglect the gravitational pull between them, the free motion 
Hamiltonian for the system would only be a sum of the individual kinetic
terms. Each particle then has the same quantum geodesic equation
governing its motion, irrespective of the actual composite state of the
two particles and to what extent they are entangled. Of course, the story
may be very different in a full treatment from a theory of quantum gravity
where one cannot simply take the metric as a fixed background. 

\subsection{Against Poincar\'e symmetry and on-shell mass condition}
So far as obtaining the geodesic equation is concerned, the Hamiltonian 
approach if the same as the Lagrangian one which is just about minimizing the
action $\int\!\! ds L$, or $\int\!\! ds \hat{L}$. One may think about it completely
independent of any dynamics. However, physics is dynamics. We have 
emphasized that from the physical point of view, the proper model of spacetime 
should be taken from the successful theory of particle dynamics rather than
assumed as given. As a parallel, the notion of a geodesic equation should only
be taken as an equation of motion for a free particle. One may want to replace
the latter with a photon. But that is really taking it beyond a theory of particle 
dynamics. In the other part of the analysis above for the WEP we have the
Hamiltonian $H_{\!a}$ with a potential from the gravitational pull as seen from 
a classical Minkowski spacetime to start with. The success of the Hamiltonian 
picture is clear. However, vigilant readers may have noticed that the on-shell
mass condition as $-p_\mu p^\mu =m_{\!\ssc E}^2 c^2$ is not necessarily
respected; it generally allows $-p_\mu p^\mu$ to have nontrivial $s$-evolution. 
Admitting a potential in violation of the on-shell mass condition is a general 
feature of the kind of fully Lorentz covariant Hamiltonian formulation \cite{H,F}. 
Note that we used a new notation $m_{\!\ssc E}$, instead of $m$ here. After all, 
the concept of Einstein rest mass is not the same as the Newtonian inertial and
gravitational mass. We have a quite elaborated discussion on the related issues
which is closely connected to the Poincar\'e symmetry that we do not see as 
the right symmetry to formulate a theory of `relativistic' or Lorentz covariant 
quantum dynamics in Ref.\cite{096}. Note that any quantum theory with 
wavefunctions as functions fo otherwise free Minkowski four-vector variables
$x^\mu$ and $\hat{p}_\mu$ as $-i(\hbar)\partial_\mu$ as in Klein-Gordon 
or Dirac equations are really not obtainable from the Poincar\'e symmetry.  
And when we have the version of the equations with the presence of an
electromagnetic interaction (through the covairant derivative), the canonical 
momentum $\hat{p}^\mu$ for the charged particle is no longer conserved. 
We do not even have $\hat{p}^\mu = m \frac{d\hat{x}^\mu}{d\tau}$. The 
same holds in the classical case. Einstein was well aware that the generally 
important quantity as the conserved momentum in any closed system may 
not be the quantity of mass times velocity and does not necessarily obey any 
on-shell mass condition \cite{E}. Note that the latter is non-negotiable for
a particle corresponding to an irreducible representation of the Poincar\'e
symmetry. Our $H_{\!a}$ with the gravitational acceleration $a$ in
a potential term is legitimate and its success in term speaks for the 
strength of the covariant Hamiltonian formulation.

\subsection{Quantum coordinate transformations and quantum gravity}
Even when we take the operator version of Eq.(\ref{cct}) and Eq.(\ref{pt}) for the 
coordinate transformation, it is really only a classical one. We talk about going
to the free-falling frame. But that is only the free-falling frame of a classical
particle. The reference frame transformation sees no quantum properties of the 
particle, and the key parameter describing the transformation, the gravitational
 acceleration $a$, is only taken as a classical quantity. To stick fully to the idea 
of the free-falling frame of the physical particle which is a quantum object, 
one should have a version of quantum reference frame transformation \cite{qrf}. 
One should take the gravitational acceleration of the quantum particle seen in 
the Minkowski frame as what it should be -- a quantum observable $\hat{a}$. 
Yet, how to do it properly is a very challenging question. To start with, going 
for a version of the operator coordinate transformation with an $\hat{a}$ has 
nontrivial operator ordering issues between $\hat{a}$ and the position 
observables $\hat{\rho}$ and $\hat\tau$ has to be resolved. 

While the idea of quantum frames of reference is not new \cite{qf1,qf2}, a more
more recent study \cite{qrf} has brought it back to the attention of many authors. 
In the example of a quantum (spatial) translation, as an explicit notion of the
position of particle $B$ relative to particle $A$, the paper gives it as a unitary 
transformation on the quantum phase space of the form 
$e^{i\hat{x}_{\!\ssc A} \hat{p}_{\ssc B}}$. The latter can be read naively as 
$\hat{p}_{\ssc B}$ generating a translation of $\hat{x}_{\ssc B}$ by `an amount'
$\hat{x}_{\!\ssc A}$ giving the position observable for particle $B$ in the
 reference frame of particle $A$ as $\hat{x}_{\!\ssc B} - \hat{x}_{\!\ssc A}$. 
Apply to explicit states of particles $A$ and $B$, interesting quantum features 
can be retrieved.  For example, if we can start with such a composite state 
that has the two positions completely entangled yet far from eigenstates of 
$\hat{x}_{\ssc A}$ and $\hat{x}_{\ssc B}$. The position of particle $B$ as 
observed from particle $A$, exactly as obtained from the transformation,
gives $B$ as in a position eigenstate. What we have is really an eigenstate of  
$\hat{x}_{\ssc B} - \hat{x}_{\ssc A}$, seen in the original frame.  One can also
starts with a product state between $A$ and $B$. From the frame of $A$, 
particle $B$ would be seen as fully entangled to the particle $C$ that 
represents the original frame of reference. Quantum properties, as such 
Heisenberg uncertainty and entanglement, generally change under 
a quantum coordinate transformation. Note the unitary transformation 
$e^{i\hat{x}_{\ssc A} \hat{p}_{\ssc B}}$ is a canonical transformation, both 
seen from the Hilbert space point of view or that of the noncommutative 
symplectic geometry of $\hat{x}$-$\hat{p}$. The transformation as given
also serves as a momentum translation  of $\hat{p}_{\ssc A}$ generated by 
the operator $\hat{x}_{\ssc A}$ by `an amount' $-\hat{p}_{\ssc B}$, hence 
maintaining the canonical condition among all pairs of position and momentum 
observables under the different frames. 

We have only given a brief sketch of the generic notion of quantum coordinate 
transformation above. Its important relevancy to a theory of quantum gravity has
been noted by Hardy\cite{Ha}. He adopts an approach to the problem quite
different from us though. Quantum gravity has to be about the quantum geometry
of the spacetime. The picture of a simple quantum spatial translation clearly
illustrates the necessity to beyond any classical geometric picture. Hardy looked 
at that quantum geometry as a superposition of classical ones. We want to look 
at it as a single noncommutative geometry, a geometry with the observables 
as coordinates. A more solid picture of the latter is given by a notion of 
noncommutative values for the observables \cite{081,079}, as a representation 
of the full quantum information a state bears for the observable. Instead of 
identifying a state as defining the evaluational functional as given by the 
expectation values of observables, it can be promoted to an evaluational
homomorphism that maps the observables to local representations of its
expectation value functions as a noncommutative algebra of the state-specific
quantum values based on the K\"ahler product of Eq.(\ref{kp}). Each such
noncommutative value can be represented by a complex number sequence
with terms essentially given by the coefficients of the Taylor series expansion
around the state. In the wavefunction picture, we have essentially the 
expectation values themselves and their functional derivatives with respect 
to the wavefunction \cite{093}. Details we refer the readers to the references. 
From the theoretical point of view, the notion of the noncommutative value for 
an observable is an element of a state-specific noncommutative algebra that 
encodes the complete mathematical information the theory of quantum 
mechanics contains when matching the observable to the state \cite{097}. 
The algebraic relations among the observables as variables are exactly 
preserved among their values. The noncommutative value for the position 
observable, for example, offers exactly an answer to what is the actual 
`amount', that value of $\hat{x}_{\ssc A}$ that the quantum spatial 
transformation discussed above translates any $\hat{x}_{\ssc B}$ with all notions 
about changes in Heisenberg uncertainty and even entanglement successfully
described \cite{093}. The noncommutative value, on the one hand, provides the
quantum generalization of the classical real number parameter of a coordinate 
transformation to make the latter quantum. On the other hand, it gives a picture 
of that noncommutative/quantum geometry as the space of all possible 
(noncommutative) values of the set of position and momentum observables 
beyond the picture of classical geometries. Definitive points, and hence 
a definite geodesic path of the quantum particle within that quantum model
of spacetime are to be specified by fixed values of the coordinate observables.
Hardy's picture of the quantum spacetime geometry is then the Schr\"odinger 
picture of our sketched Heisenberg picture here. 

It is interesting to note, in relation to our discussion, a comment from Penrose 
n the problem of compatibility of
quantum mechanics and the principle of relativity \cite{P}. Basically, Penrose
was pointing to the fact that no reference frame transformation can take
a position eigenstate to another state that is not a position eigenstate. But
that is true only when the classical picture of all possible eigenvalues of
the position observable is taken to give the model of that space(time).
With the notion of quantum reference frame transformation, one can apply
a translation by exactly the (noncommutative value) amount of difference
in the definite noncommutative values of the two positions to take one to
the other. A point in the noncommutative space(time) can be described by
the different noncommutative coordinate values under different choices
of reference frames. There is no intrinsic difference between a point with
a noncommutative position coordinate value that corresponds to an
eigenstate and one that has a more nontrivial noncommutative position 
coordinate value. This illustrates well then quantum general relativity is
about quantum reference frame transformations that certainly cannot
be formulated in terms of a classical space(time) manifold.

\section*{Acknowledgement}
The author thanks P.-M. Ho and J.-T. Hsiang for discussions. Thanks also go
to B.L. Hu for reading the manuscript, and comments.
The work is partially supported by research grant 
number    112-2112-M-008-019 of the NSTC of Taiwan.

\end{document}